\documentclass[12pt,a4paper]{article}
\usepackage[dvips]{graphicx}
\usepackage{amssymb}
\usepackage[centerlast,footnotesize]{caption2}
\usepackage{float}

\usepackage{graphicx}
\newcommand{\One}{1\kern-4.5pt1}

\newcommand{\be}{\begin{equation}}
\newcommand{\ee}{\end{equation}}

\def\lesim{${\lower 2pt\hbox{$\scriptstyle
<$}\atop\raise 4pt\hbox{$\scriptstyle\sim$}}$} 
\def\grsim{${\lower2pt\hbox{$\scriptstyle >$} \atop\raise4pt\hbox 
{$\scriptstyle\sim$}}$} 

\begin{document}
\begin{center}
\begin{flushright}
December 2015
\end{flushright}
\vskip 10mm
{\LARGE
From Domain Wall to Overlap in 2+1$d$
}
\vskip 0.3 cm
{\bf Simon Hands}
\vskip 0.3 cm
{\em Department of Physics, College of Science, Swansea University,\\
Singleton Park, Swansea SA2 8PP, United Kingdom.}
\end{center}

\noindent
{\bf Abstract:} 
The equivalence of domain wall and overlap fermion formulations is demonstrated for
lattice gauge theories in 2+1 spacetime dimensions with parity-invariant mass
terms. Even though the domain wall approach distinguishes propagation along a
third direction with projectors ${1\over2}(1\pm\gamma_3)$, the truncated
overlap operator obtained for finite wall separation $L_s$ is
invariant under interchange of $\gamma_3$ and $\gamma_5$. In the limit
$L_s\to\infty$ the resulting Ginsparg-Wilson relations recover the expected
U(2$N_f$) global symmetry up to O($a$) corrections. Finally it is shown that
finite-$L_s$ corrections to bilinear condensates associated with dynamical mass
generation are characterised by whether
even powers of the symmetry-breaking mass are present; such terms are absent for
antihermitian bilinears such as $i\bar\psi\gamma_3\psi$, 
markedly improving the approach to the large-$L_s$
limit.
\vspace{0.5cm}

\noindent
Keywords: 
Lattice Gauge Field Theories, Field Theories in Lower Dimensions, Global
Symmetries

\section{Introduction}

Relativistic fermions moving in 2 spatial dimensions are the focus of much
attention, in part due to the stability of Dirac points in graphene and surface
states of topological band insulators when the underlying Hamiltonian is symmetric
under time reversal and spatial inversion (see, eg. \cite{Bernevig}). Even in this case 
a gap may develop at the Dirac points in the presence of interactions. The 
corresponding issue in quantum field theory is the stability of the vacuum 
with respect to spontaneous generation of a parity-invariant bilinear condensate of the
form $\langle\bar\psi\Gamma_i\psi\rangle\not=0$. 
Since the transition to a gapped phase generically occurs for strong interactions,
it defines a quantum critical point (QCP)~\cite{Son:2007ja}; the phase diagram 
for planar fermionic systems with various interactions and characterisation of
possible QCPs as a function of the number of fermion species $N_f$
remain open questions~\cite{Janssen:2012pq}.

To date there have been many lattice field theory simulations probing QCPs using
the staggered fermion formulation~\cite{DelDebbio:1997dv} (a notable recent exception employs the SLAC
derivative~\cite{Schmidt:2015fps}); $N$ staggered fermions describe $N_f=2N$ continuum flavors each
having 4 spinor components~\cite{Burden:1986by}, with
global symmetry group U($N)\otimes$U($N$) spontaneously broken by a parity-invariant mass to
U($N$). However, because there are two matrices $\gamma_3$ and $\gamma_5$ which
anticommute with the kinetic operator, the correct continuum symmetry breaking
is U($2N_f)\to$U($N_f)\otimes$U($N_f$). For the strongly-interacting continuum
limit
at a QCP, there is no reason {\it a priori\/} to expect the correct
symmetry-breaking pattern to be recovered.

For this reason the properties of domain wall fermions, which purportedly more
faithfully
reproduce continuum symmetries, were explored for 2+1+1$d$ in
Ref.~\cite{Hands:2015qha}. In particular bilinear condensates and meson
correlators constructed from distinct spinor combinations, but which
should yield identical results in a U(2)-invariant theory, were investigated as
a function of the extent $L_s$ of the ``third'' direction separating the domain
walls. 
Numerical results obtained in the context of quenched non-compact QED$_3$ with
variable coupling strength support U(2) symmetry being restored as
$L_s\to\infty$.  In 2+1$d$ the Ginsparg-Wilson relation specifying the optimal
requirements for lattice fermions to avoid species doubling while retaining as
much of the continuum global symmetry as possible~\cite{Ginsparg:1981bj} generalises to a set of three
relations (since chiral rotations are now specified by an element of U(2) rather
than U(1)). These were set out in \cite{Hands:2015qha},
along with the specification of an overlap Dirac operator
$D_{ov}$~\cite{Neuberger:1997fp} defined in 2+1$d$ in which 
realises them.  As it must, $D_{ov}$ has equivalent properties under
the U(2) rotations generated by $\gamma_3$ and $\gamma_5$. 

In the domain wall approach, the 2+1$d$ fields $\psi,\bar\psi$ are
defined in terms of surface states 
fields $\Psi_\pm,\bar\Psi_\pm$ 
which are approximately localised on the walls and are $\pm$ eigenstates of 
$\gamma_3$~\cite{Kaplan:1992bt}. 
Some questions which remain unanswered in \cite{Hands:2015qha} are: the extent
to which the
domain wall formulation, in which propagation along the direction separating the
walls is governed by $\gamma_3$, can maintain the equivalence between
$\gamma_3$ and $\gamma_5$ rotations for finite $L_s$; the reason for $O(a)$
violations of U(2) symmetries even in the overlap limit $L_s\to\infty$;
and a better understanding of why finite-$L_s$ corrections are minimised by
choosing $i\langle\bar\psi\gamma_3\psi\rangle$, rather than
$\langle\bar\psi\psi\rangle$, as the bilinear condensate to focus on.
In this brief technical Letter I outline how the overlap operator is recovered in
the $L_s\to\infty$ limit of the domain wall formulation using a by now familiar
sequence of matrix algebra operations. In particular, it will prove possible to
extend the key results on the equivalence of $\gamma_3$ and $\gamma_5$ 
to a truncated overlap operator defined by domain wall fermions
with finite $L_s$. 
As well as providing a firm conceptual
foundation for domain wall fermions and their symmetry properties in 2+1$d$, the
proof sheds light on each of these outstanding issues.

\section{From Domain Wall to Overlap}
\label{sec:DW2ov}

First we review the passage from the domain wall formulation of lattice fermions
to the overlap operator. The corresponding treatment for 4$d$ gauge theories is
well-known~\cite{Neuberger:1997bg}: here we
follow closely the treatment of \cite{Kennedy:2006ax}.
We begin from the $2+1d$ domain wall operator defined in~\cite{Hands:2015qha},
correcting an overall (unphysical) sign:
\begin{equation}
S^{dw}=\sum_{x,y}\sum_{s,r}\bar\Psi(x,s)D(x,s\vert y,r)\Psi(y,r),
\label{eq:SDWF}
\end{equation}
The fields $\Psi,\bar\Psi$ are four-component spinors defined in 2+1+1
dimensions, 
and 
\begin{equation}
D(x,s\vert y,s^\prime)=\delta_{s,r}D_W(x\vert y)+\delta_{x,y}D_3(s\vert s^\prime),
\label{eq:DW}
\end{equation}
where the first term is the $2+1d$ Wilson operator defined on spacetime volume
$V$
\begin{equation}
(D_W-M)_{x,y}=-{1\over2}\sum_{\mu=0,1,2}
\left[(1-\gamma_\mu)U_\mu(x)\delta_{x+\hat\mu,y}+(1+\gamma_\mu)U^\dagger_\mu(y)\delta_{x-\hat\mu,y}
\right]
+(3-M)\delta_{x,y},
\label{eq:Ddw}
\end{equation}
and $D_3$ controls hopping along the dimension separating the domain walls at
$s=1$ and $s=L_s$, which
we will refer to as the third direction:
\begin{equation}
D_{3\,s,s^\prime}
=-\left[P_-\delta_{s+1,s^\prime}
(1-\delta_{s^\prime,L_s})
+P_+\delta_{s-1,s^\prime}(1-\delta_{s^\prime,1})\right]
+\delta_{s,s^\prime},
\label{eq:D3dw}
\end{equation}
where the projectors $P_\pm\equiv{1\over2}(1\pm\gamma_3)$.
Following convention, in (\ref{eq:Ddw}) we include interaction with a SU($N_c$)
valued gauge connection 
field $U_\mu(x)$ located on the lattice links, noting in passing that some 
models relevant for 2+1$d$ QCPs share the global U(2$N_f$) symmetries of gauge theories. 

Initially we supplement (\ref{eq:SDWF}) with a hermitian mass term coupling fields
on opposite walls:
\begin{equation}
m_hS_h=m_h\sum_x\bar\Psi(x,L_s)P_-\Psi(x,1)+\bar\Psi(x,1)P_+\Psi(x,L_s).
\label{eq:m_h}
\end{equation}

The operator $D_W-M+D_3+m_hS_h$ can be represented as a $L_s\times L_s$ matrix
consisting of $4VN_c\times4VN_c$ blocks:
\begin{eqnarray}
D(m_h)&=&
\left[
\begin{matrix}{
D_W-M+1 & 0 & \cdots& &+m_h\cr
-1 & D_W-M+1 & 0 & &\cr
0 & -1 & \ddots & &\cr
\vdots& & & & \cr
0& & &-1 & D_W-M+1\cr}
\end{matrix}
\right]P_+\nonumber\\&+&
\left[
\begin{matrix}{
D_W-M+1 &  -1 & \cdots & &0\cr
0 & D_W-M+1 & -1 & &\cr
\vdots& 0& \ddots & &\cr
& & & & -1\cr
+m_h & & && D_W-M+1\cr}
\end{matrix}
\right]P_-.\label{eq:DWmatrix}
\end{eqnarray}
Now define the cyclical shift operator ${\cal
P}_{s,s^\prime}
\equiv[\delta_{s-1,s^\prime}(1-\delta_{s,1})+\delta_{s,1}\delta_{s^\prime,L_s}]P_-+\delta_{s,s^\prime}P_+$
so that
\begin{equation}
D{\cal P}=\left[\begin{matrix}{
Q_+&0&\cdots&&Q_-C_-\cr
Q_-&Q_+&0&&0\cr
0&Q_-&\ddots&\ddots&\vdots\cr
\vdots&\ddots&\ddots&&0\cr
0&&&Q_-&Q_+C_+\cr}
\end{matrix}\right]
\label{eq:DP}
\end{equation}
with 
\begin{eqnarray}
Q_\pm&=&(D_W-M+1)P_\pm-P_\mp;\\
C_\pm(m_h)&=&{1\over2}(1-m_h)\pm{1\over2}(1+m_h)\gamma_3=P_\pm-m_hP_\mp.\label{eq:Cpm}
\end{eqnarray}
Now define the block diagonal matrix ${\cal Q}=Q_+\One_{4VN_c\times4VN_c}$; it
is important to note that $Q_\pm\not=Q_\pm(m_h)$, ${\cal Q}\not={\cal Q}(m_h)$,
${\cal P}\not={\cal P}(m_h)$.
With $\tilde
D\equiv{\cal Q}^{-1}D{\cal P}$, we deduce
\begin{equation}
\mbox{det}[\tilde D(1)^{-1}\tilde D(m_h)]\equiv\mbox{det}[D(1)^{-1}D(m_h)],
\end{equation}
where
\begin{equation}
\tilde D=\left[\begin{matrix}{
1&0&\cdots&&-T^{-1}C_-\cr
-T^{-1}&1&0&&0\cr
0&-T^{-1}&1&&\vdots\cr
\vdots&&\ddots&\ddots&0\cr
0&&&-T^{-1}&C_+\cr}
\end{matrix}
\right],\label{eq:ECT}
\end{equation}
with $T=-Q_-^{-1}Q_+$.

In more detail,
\begin{eqnarray}
T&=&-[(D_W-M+1)P_--P_+]^{-1}[(D_W-M+1)P_+-P_-]\nonumber\\
&=&\left[1-\gamma_3{{(D_W-M)}\over{2+(D_W-M)}}\right]^{-1}
\left[1+\gamma_3{{(D_W-M)}\over{2+(D_W-M)}}\right]
={{1-H}\over{1+H}}
\end{eqnarray}
where the hermitian $4VN_c\times4VN_c$ matrix $H$ is defined
\begin{equation}
H=-\gamma_3[2+(D_W-M)]^{-1}[D_W-M]\equiv-\gamma_3A.
\label{eq:H}
\end{equation}
Hermiticity of $H$ requires $\gamma_3A\gamma_3=A^\dagger$, which is the case for
$A$ defined by (\ref{eq:Ddw}).
Up to an unphysical sign and with $\gamma_3$ assuming the role played by
$\gamma_5$ in 4$d$ gauge theories, $H$ is
identical with the Shamir kernel~\cite{Shamir:1993zy}.

Next observe that in the form (\ref{eq:ECT}), $\tilde D=L{\cal D}U$ with
\begin{equation}
L=\left[\begin{matrix}{
1&0&\cdots&&0\cr
-T^{_1}&1&0&&\vdots\cr
0&-T^{-1}&\ddots&&\cr
\vdots&&\ddots&&\cr
0&&&-T^{-1}&1\cr}
\end{matrix}\right];\;\; 
U=\left[\begin{matrix}{
1&0&\cdots&&-T^{-1}C_-\cr
0&1&0&&-(T^{-1})^2C_-\cr
\vdots&&1&\ddots&-(T^{-1})^3C_-\cr
&&&\ddots&\vdots\cr
0&&&&1\cr}
\end{matrix}\right].
\end{equation}
and
\begin{equation}
{\cal D}=\left[\begin{matrix}{
1&0&\cdots&&0\cr
0&1&0&&\vdots\cr
\vdots&&1&&\cr
&&&\ddots&\cr
0&&&&C_+-(T^{-1})^{L_s}C_-\cr}
\end{matrix}
\right],
\end{equation}
Again, note $L\not=L(m_h)$, and $\mbox{det}L=\mbox{det}U=1$.
We conclude 
\begin{equation}
\mbox{det}[D(1)^{-1}D(m_h)]
=\mbox{det}[\tilde D(1)^{-1}
\tilde D(m_h)]=\mbox{det}[{\cal D}_{L_s,L_s}(1)^{-1}{\cal D}_{L_s,L_s}(m_h)], 
\end{equation}
where the $4VN_c\times4VN_c$ matrix ${\cal D}_{L_s,L_s}$ is the Schur complement of $\tilde
D$:
\begin{eqnarray}
{\cal D}_{L_s,L_s}(m_h)&=&C_+-(T^{-1})^{L_s}C_-
=(1+{\cal T}^{-1})\gamma_3{1\over2}\left[(1+m_h)-(1-m_h)\gamma_3{{1-{\cal
T}}\over{1+{\cal T}}}\right]\nonumber\\
&=&{\cal D}_{L_s,L_s}(1){1\over2}\left[(1+m_h)-(1-m_h)\gamma_3{{1-{\cal
T}}\over{1+{\cal T}}}\right],\label{eq:Schur}
\end{eqnarray}
with ${\cal T}\equiv T^{L_s}$. We now multiply both sides of (\ref{eq:Schur})
by $D_{L_s,L_s}^{-1}(1)$ to find that the combination of domain wall fermion
determinants $\mbox{det}[D(1)^{-1}D(m_h)]$ is the same as the determinant of the
{\it truncated overlap\/} operator
\begin{eqnarray}
D_{Ls}[H]&=&{1\over2}\left[(1+m_h)-(1-m_h)\gamma_3{{1-\left({{1-H}\over{1+H}}\right)^{L_s}}\over
{1+\left({{1-H}\over{1+H}}\right)^{L_s}}}\right]\\
&\equiv&{1\over2}\left[(1+m_h)-(1-m_h)\gamma_3\tanh(L_s\tanh^{-1}H)\right].
\label{eq:tanh}
\end{eqnarray}
In order for the tanh function to be defined by a power series the 
second equality (\ref{eq:tanh}) requires $H$ to be a bounded operator,
namely $\vert H\vert<1$.
The factor $D(1)^{-1}$ can be thought of as modelling Pauli-Villars boson
fields which cancel the contributions of the fermions from the 4$d$
bulk.
Now, $\tanh(L_s\tanh^{-1}(x))$ is an analytic approximation to the signum
function $\mbox{sgn}(x)$ which becomes exact in the limit $L_s\to\infty$.
So long as $H$ is hermitian and bounded, we therefore recover the overlap
operator~\cite{Neuberger:1997fp}:
\begin{eqnarray}
\lim_{L_s\to\infty}D_{L_s}=D_{ov}&=&{1\over2}\left[(1+m_h)-(1-m_h)\gamma_3\mbox{sgn}
\left(-\gamma_3{{D_W-M}\over{2+(D_W-M)}}\right)\right]\nonumber\\
&=&{1\over2}\left[(1+m_h)+(1-m_h){{A}\over\sqrt{A^\dagger A}}\right],\label{eq:overlap}
\end{eqnarray}
where the unphysical nature of the sign of $\gamma_3$ is
manifest. For $m_h\to0$ (\ref{eq:overlap}) coincides with the 2+1$d$ overlap operator given
in \cite{Hands:2015qha}.

Next let's check the overlap operator (\ref{eq:overlap}) has the expected
weak-coupling limit. For link fields $U_\mu=1$, and with lattice spacing set to
unity,  
in momentum space $D_W=i\sum_\mu\gamma_\mu\sin p_\mu+\sum_\mu(1-\cos p_\mu
)$, implying propagator poles at $p_\mu\approx0$ and near the Brillouin Zone
corners $p_\mu\approx\pi$. At the origin $D_W\approx i\gamma_\mu p_\mu$
so 
\begin{equation}
\mbox{sgn}(H)={H\over\sqrt{H^2}}\approx-\gamma_3{{(ip{\!\!\!
/}-M)}\over{(2-M)}}{(2-M)\over M}=-\gamma_3\left[{{ip{\!\!\! /}}\over M}-1\right]
\label{eq:wc}
\end{equation}
so that the overlap operator 
\begin{equation}
D_{ov}\approx ip{\!\!\! /}{{(1-m_h)}\over 2M}+m_h.
\end{equation}
Taking into account a benign wavefunction renormalisation, this is the propagator for a
continuum species with mass proportional to $m_h$. By contrast
near a doubler pole $\tilde p_\mu=p_\mu-(i,j,k)\pi\approx0$, $i,j,k\in\{+1,-1\}$,
\begin{equation}
\mbox{sgn}(H)\approx-\gamma_3{{i\tilde p{\!\!\!
/}+(2n-M)}\over{(2n-M)}}=-\gamma_3\left[{{i\tilde p{\!\!\! /}}\over (2n-M)}+1\right]
\end{equation}
with $n=\vert i\vert+\vert j\vert+\vert k\vert$, so the overlap is
\begin{equation}
D_{ov}\approx1+ {{(1-m_h)}\over{2(2n-M)}}i\tilde p{\!\!\! /}.
\end{equation}
So long as $(2n-M)$ is not too small, the species has a mass of O(1) in cutoff units, 
and 
decouples from low-energy physics.

\section{Equivalence of $\gamma_3$ and $\gamma_5$}
\label{sec:325}

Despite the manifest independence of the overlap operator $D_{ov}$
(\ref{eq:overlap}) of which matrix $\gamma_3$ or $\gamma_5$ is used to
define the hermitian argument $H$ of the signum function, for finite $L_s$ it
remains unclear whether the distinction is important or
not~\cite{Hands:2015qha}, since clearly the definition (\ref{eq:D3dw}) of the domain wall
operator $D_3$ distinguishes them. We can address this using the analytic approximation
for signum (\ref{eq:tanh}). 

First, the series expansion for $\tanh^{-1}H$ is well-defined since $H=\gamma_3A$ is
a bounded operator, ie. $\vert H\vert=M/(2-M)<1$  for $0<M<1$
\footnote{
For free fermions the most stringent limit on $M$ comes from the origin of momentum
space.
In practice on any finite lattice with antiperiodic temporal boundary conditions
$M=1$ is safe since
$\vert H\vert
= 1/\sqrt{5-4\cos{\pi\over L_t}}<1$
for $L_t<\infty$.}:
\begin{equation}
\tanh^{-1}H=H+{H^3\over3}+{H^5\over5}+\cdots
\end{equation}
Each term is on odd power, so can be reexpressed using
$\gamma_3A\gamma_3=A^\dagger$:
\begin{equation}
H^{2n+1}=\gamma_3A(A^\dagger A)^n.
\end{equation}
The signum approximation is then
\begin{equation}
\tanh(L_s\gamma_3A\sum_n b_n(A^\dagger A)^n)
={{\sinh(L_s\gamma_3A\sum_n b_n(A^\dagger A)^n)}\over
{\cosh(L_s\gamma_3A\sum_n b_n(A^\dagger A)^n)}}
\label{eq:sinhcosh}
\end{equation}
with $b_n=(2n+1)^{-1}$. In the McLaurin series expansions of the
hyperbolic functions on the RHS of (\ref{eq:sinhcosh}), expansion of the
argument yields a general term of the form 
\begin{equation}
L_s^m\left(\sum_{n_1=0}^\infty\sum_{n_2=0}^\infty\cdots\sum_{n_{m}}^\infty\right)
\prod_{i=1}^m[b_{n_i}(\gamma_3A)(A^\dagger A)^{n_i}]
\end{equation}
For the sinh series, $m$ is an odd integer so that the term in square brackets reads
\begin{eqnarray}
(\prod b_{n_i})(\gamma_3A)(A^\dagger)^{n_1}(\gamma_3A)(A^\dagger
A)^{n_2}\ldots(\gamma_3A)(A^\dagger A)^{n_m}&=&\nonumber\\
(\prod b_{n_i})(\gamma_3A)(A^\dagger)^{n_1}(A^\dagger
A)^{n_2+1}(A^\dagger A)^{n_3}\ldots(A^\dagger A)^{n_{m-1}+1}(A^\dagger A)^{n_m}&=&\nonumber\\
(\prod b_{n_i})(\gamma_3A)(A^\dagger A)^{\sum_i n_i+(m-1)/2}.
\end{eqnarray}
For the cosh series $m$ is even and a similar argument gives the general
term\hfill\break
$(\prod b_{n_i})(A^\dagger A)^{\sum_i n_i+m/2}$. 

The final step is to observe
that $[(\gamma_3A)^{-1},(A^\dagger A)^n]=0$ for any $n$; the RHS of (\ref{eq:sinhcosh})
can therefore be manipulated to bring $\gamma_3A$ to the left of all terms in
the expansion, whereupon the $\gamma_3$ cancels in the expression (\ref{eq:tanh})
for the truncated overlap. Now using the fact that $\gamma_5$ has identical
properties with respect to commutation with $A$, we can reverse
all the steps to rewrite the truncated overlap operator
\begin{equation}
D_{Ls}[H]
={1\over2}\left[(1+m_h)+(1-m_h)\gamma_5\tanh(L_s\tanh^{-1}\gamma_5A)\right].
\label{eq:tanh5}
\end{equation}
This establishes that the truncated overlap operator is equally blind to the
distinction between $\gamma_3$ and $\gamma_5$ as the overlap (\ref{eq:overlap}).

\section{Introducing $m_3,m_5\not=0$}
\label{sec:m3m5}

In \cite{Hands:2015qha} we exploited the possibility of U(2)-rotating the fields
leaving the kinetic term unaltered while changing the form of the mass term. In
terms of continuum fields defined in 2+1$d$ the alternative but physically
equivalent,
antihermitian but parity-invariant mass terms are $im_3\bar\psi\gamma_3\psi$,
$im_5\bar\psi\gamma_5\psi$. In the domain wall approach (\ref{eq:m_h}) is
replaced by one of
\begin{eqnarray}
m_3S_3&=&im_3\sum_x\bar\Psi(x,L_s)\gamma_3P_-\Psi(x,1)+\bar\Psi(x,1)\gamma_3P_+\Psi(x,L_s);
\label{eq:m_3}\\
m_5S_5&=&im_5\sum_x\bar\Psi(x,L_s)\gamma_5P_+\Psi(x,L_s)+\bar\Psi(x,1)\gamma_5P_-\Psi(x,1).
\label{eq:m_5}
\end{eqnarray}

First consider a mass term $m_3S_3$. The matrix manipulations outlined in
Sec.~\ref{sec:DW2ov} leading to eqn.~(\ref{eq:DP}) go through as before, 
but with (\ref{eq:Cpm}) replaced by
\begin{equation}
C_{3\pm}=P_\pm\pm im_3P_\mp,
\end{equation}
The Schur complement of $\tilde D={\cal Q}^{-1}D{\cal P}$ is then
\begin{equation}
{\cal D}_{L_s,L_s}(m_h=1){1\over2}\left[
(1+im_3\gamma_3)-\gamma_3{{1-{\cal T}}\over{1+{\cal
T}}}(1-im_3\gamma_3)\right],
\end{equation}
implying a truncated overlap
\begin{equation}
D_{L_s}={1\over2}\left[(1+im_3\gamma_3)-\gamma_3\tanh(L_s\tanh^{-1}(\gamma_3A))(1-im_3\gamma_3)\right],
\label{eq:g3}
\end{equation}
with $A$ still given by (\ref{eq:H}). An important technical point is that the passage
from domain wall to overlap requires the Pauli-Villars matrix ${\cal
D}_{L_s,L_s}(1)=(1+{\cal T}^{-1})\gamma_3$ to continue to be defined with the
{\it hermitian\/} mass term $1\times S_h$.
The overlap operator found in the limit $L_s\to\infty$ is thus
\begin{equation}
D_{ov}={1\over2}\left[(1+im_3\gamma_3)+{A\over{\sqrt{A^\dagger
A}}}(1-im_3\gamma_3)\right]\label{eq:overlap3}
\end{equation} 
with $A$ defined in (\ref{eq:H}).  In the weak coupling long wavelength limit
\begin{equation}
D_{ov}\approx ip{\!\!\! /}{{(1-im_3\gamma_3)}\over{2M}}+im_3\gamma_3.
\end{equation}
This time there is an $O(a)$  term proportional to $p{\!\!\!/}\gamma_3$ not present in
the continuum action, which cannot be absorbed by wavefunction rescaling. 
It seems highly plausible that
this lies at the heart of the $O(a)$ departures from U(2) symmetry observed when
rotating fermion bilinears according to the remnant symmetries derived from
the 3$d$ Ginsparg-Wilson (GW) relations in Sec.~3 of \cite{Hands:2015qha}.

Next consider the mass term $m_5S_5$. Even though this term differs from
the other masses by coupling fields on the same domain wall to itself,
rather than on opposite ones,
the matrix manipulations of Sec.~\ref{sec:DW2ov} still arrive at
(\ref{eq:DP}), with this time
\begin{equation}
C_{5\pm}=P_\pm-im_5\gamma_5P_\pm=P_\pm-im_5P_\mp\gamma_5,
\end{equation}
where the second step is crucial. The truncated overlap in this case is
\begin{equation}
D_{L_s}[H]={1\over2}\left[(1+im_5\gamma_5)-\gamma_3\tanh(L_s\tanh^{-1}H)(1-im_5\gamma_5)\right];
\end{equation}
however the considerations of Sec.~\ref{sec:325} permit this to be rewritten
\begin{equation}
D_{L_s}={1\over2}\left[(1+im_5\gamma_5)-\gamma_5\tanh(L_s\tanh^{-1}(\gamma_5A))(1-im_5\gamma_5)\right].
\label{eq:g5}
\end{equation}
The complete equivalence between (\ref{eq:g5}) and (\ref{eq:g3}) is manifest.

\section{Ginsparg-Wilson Relations}
\label{sec:GW}

Whilst the previous two sections have established the equivalence of the domain
wall formulation with respect to a discrete interchange of the matrices $\gamma_3$ and
$\gamma_5$, in order to study restoration of the full
U($2N_f$) symmetry it is more convenient to examine the 
overlap operator. 
Following (\ref{eq:overlap},\ref{eq:overlap3}), in the large-$L_s$ limit we can write Lagrangian
densities in terms of ``Ginsparg-Wilson'' fields $\Psi,\bar\Psi$:
\begin{eqnarray}
{\cal L}_h&=&\bar\Psi[D_{ov}^0+m_h(1-D_{ov}^0)]\Psi;\label{eq:Lh}\\
{\cal L}_3&=&\bar\Psi[D_{ov}^0+im_3(1-D_{ov}^0)\gamma_3]\Psi,\label{eq:L3}
\end{eqnarray}
where $D_{ov}^0$ is the overlap operator for massless fermions
\begin{equation}
D_{ov}^0={1\over2}\left[1+{A\over\sqrt{A^\dagger A}}\right].
\label{eq:D0}
\end{equation}
In both cases there is an $O(a)$ correction to the expected continuum form, but
as noted above for the hermitian mass case (\ref{eq:Lh}) the correction can be absorbed into a
harmless rescaling of the kinetic term. For the antihermitian case (\ref{eq:L3}) by contrast
the correction is not of the same form as a term in the continuum Lagrangian, as
first noted in \cite{Hands:2015qha} (although (\ref{eq:L3}) differs in detail
from eq.~(34) of that paper).

The reconciliation is made by first observing that the GW relation appropriate
for the domain wall operator (\ref{eq:DW}) is~\cite{Ginsparg:1981bj,Kennedy:2006ax}
\begin{equation}
\gamma_3D_{ov}^0+D_{ov}^0\gamma_3=2D_{ov}^0\gamma_3D_{ov}^0.
\label{eq:GW}
\end{equation}
As expected, there are further GW relations, first  with $\gamma_5$ replacing
$\gamma_3$ in (\ref{eq:GW}), and also a rotation generated by
$i\gamma_3\gamma_5$ which along with a simple global phase rotation 
completely specifies the U(2):
\cite{Hands:2015qha}: 
\begin{equation}
\gamma_5D_{ov}^0+D_{ov}^0\gamma_5=2D_{ov}^0\gamma_5D_{ov}^0;\;\;
\gamma_3\gamma_5D_{ov}^0-D_{ov}^0\gamma_3\gamma_5=0.
\label{eq:GW535}
\end{equation}
The associated symmetry in the massless limit is
then~\cite{Luscher:1998pqa,Hands:2015qha}
\begin{eqnarray}
\Psi\mapsto e^{(i\alpha\gamma_3(1-D_{ov}^0))}\Psi&;&\;\;\;
\bar\Psi\mapsto \bar\Psi e^{(i\alpha(1-D_{ov}^0)\gamma_3)}\nonumber\\
\Psi\mapsto e^{(i\alpha\gamma_5(1-D_{ov}^0))}\Psi&;&\;\;\;
\bar\Psi\mapsto \bar\Psi e^{(i\alpha(1-D_{ov}^0)\gamma_5)}\\
\Psi\mapsto e^{-\alpha\gamma_3\gamma_5}\Psi&;&\;\;
\bar\Psi\mapsto\bar\Psi e^{\alpha\gamma_3\gamma_5}.\nonumber
\end{eqnarray}
Strictly speaking, therefore, symmetry under global U(2) rotations of local
fields is only recovered as $a\to0$, under the assumption
that the overlap operator $D_{ov}^0$ is sufficiently localised in this limit.

Next, define projection operators as follows: 
\begin{equation}
P_\pm={1\over2}(1\pm\gamma_3);\;\;\;\tilde P_\pm=P_\pm\mp D_{ov}^0\gamma_3
\end{equation}
with the property $\tilde P_\pm D_{ov}^0=D_{ov}^0P_\mp$ following from
(\ref{eq:GW}). With projected fields $\Psi_\pm=P_\pm\Psi$,
$\bar\Psi_\pm=\bar\Psi\tilde P_\mp$, we can write
\begin{eqnarray}
{\cal L}^0&=&\bar\Psi_+D_{ov}^0\Psi_++\bar\Psi_-D_{ov}^0\Psi_-=\bar\Psi
D_{ov}^0\Psi;\\
m_hS_h^{GW}&=&m_h(\bar\Psi_-\Psi_++\bar\Psi_+\Psi_-)=m_h\bar\Psi(1-D_{ov}^0)\Psi;\\
m_3S_3^{GW}&=&im_3(\bar\Psi_-\gamma_3\Psi_++\bar\Psi_+\gamma_3\Psi_-)=
im_3\bar\Psi(1-D_{ov}^0)\gamma_3\Psi\label{eq:m3S3}
\end{eqnarray}
consistent with (\ref{eq:Lh},\ref{eq:L3}). The extension to the terms involving
$\gamma_5$ is trival~\cite{Hands:2015qha}.
\footnote{
Note that in order to recover the expressions
for the antihermitian mass terms 
derived in \cite{Hands:2015qha} we should have chosen a matrix decomposition of
$D(m_{3,5})$
with the projectors $P_\pm$ multiplying to the left
rather than to the right as in (\ref{eq:DWmatrix}).}

\section{Bilinear Condensates}
\label{sec:bilcon}

The freedom to specify variants of the
parity-inavriant mass term can be exploited in the study of the corresponding bilinear
condensates defined via
\begin{equation}
\langle\bar\psi\Gamma_i\psi\rangle=
{{\partial\ln{\cal Z}}\over{\partial m_i}}=\left\langle\mbox{tr}M^{-1}{{\partial
M}\over{\partial m_i}}\right\rangle,
\end{equation}
where $\mbox{det}M$ is the part of the functional measure coming from the
fermions. For a U(2)-invariant theory the condensates generated by the masses
$m_h,m_3,m_5$ should all coincide, and indeed numerical evidence for this as
$L_s\to\infty$ was presented for quenched non-compact
QED$_3$~\cite{Hands:2015qha}. A particular useful result was that finite-$L_s$
corrections are minimised by choosing the mass term antihermitian. We
parametrise these in terms of residuals
$\Delta_h,\epsilon_h,\epsilon_3,\epsilon_5$ which vanish exponentially as
$L_s\to\infty$ by writing:
\begin{eqnarray}
{1\over2}\langle\bar\psi\psi\rangle_{L_s}&=&{i\over2}\langle\bar\psi\gamma_3\psi\rangle_{L_S\to\infty}
+\Delta_h(L_s)+\epsilon_h(L_s);\nonumber\\
{i\over2}\langle\bar\psi\gamma_3\psi\rangle_{L_s}&=&{i\over2}\langle\bar\psi\gamma_3\psi\rangle_{L_S\to\infty}
+\epsilon_3(L_s);\label{eq:residuals}\\
{i\over2}\langle\bar\psi\gamma_5\psi\rangle_{L_s}&=&{i\over2}\langle\bar\psi\gamma_3\psi\rangle_{L_S\to\infty}
+\epsilon_5(L_s).\nonumber
\end{eqnarray}
The numerically dominant residual is $\Delta_h$, defined to be the
imaginary component of $i\langle\bar\psi\gamma_3\psi\rangle$ evaluated on just
the $+$ component of $\Psi$:
\begin{equation}
i\langle\bar\Psi(1)\gamma_3P_+\Psi(L_s)\rangle={i\over2}\langle\bar\psi\gamma_3\psi\rangle_{L_s}+i\Delta_h(L_s).
\end{equation}
The imaginary contribution from the $\Psi_-$ component has opposite sign and
hence cancels even for finite $L_s$.

In order to understand why $\Delta_h$ only contributes for the hermitian
condensate, 
first consider the continuum case with $M=D{\!\!\!\! /}\,+m_h$:
\begin{eqnarray}
\langle\bar\psi\psi\rangle=\mbox{tr}(D{\!\!\!\! /}\,+m_h)^{-1}&=&\mbox{tr}
{1\over{D{\!\!\!\! /}}}\left[1-{m_h\over{D{\!\!\!\! /}}}+{m_h^2\over{D{\!\!\!\!
/}^{\,2}}}-{m_h^3\over{D{\!\!\!\! /}^{\,3}}}+\cdots\right]\nonumber\\
&=&-4\left[{m_h\over D^2}+{m_h^3\over D^4}+\cdots\right]
\end{eqnarray}
where we assume $m_h$ is small enough to justify the binomial expansion. Since
the trace over an odd number of gamma matrices is zero, all
even powers of $m_h$ vanish on taking the trace, which makes sense since
$\langle\bar\psi\psi\rangle$ should be an odd function of $m_h$. The mass term 
$m_3S_3$ yields the same series:
\begin{eqnarray}
i\langle\bar\psi\gamma_3\psi\rangle=\mbox{tr}(D{\!\!\!\!
/}\,+im_3\gamma_3)^{-1}i\gamma_3&=&\mbox{tr}
\left[1-{im_3\over{D{\!\!\!\! /}}}\gamma_3+{m_3^2\over{D{\!\!\!\!
/}^{\,2}}}-{im_h^3\over{D{\!\!\!\! /}^{\,3}}}\gamma_3+\cdots\right]
{1\over{D{\!\!\!\! /}}}(i\gamma_3)\nonumber\\
&=&-4\left[{m_3\over D^2}+{m_3^3\over D^4}+\cdots\right]
\end{eqnarray}
where we have used $\gamma_3D{\!\!\!\! /}\gamma_3=-D{\!\!\!\! /}$. This time the
even powers vanish because they consist of products of an odd number of
matrices $\gamma_\mu$  ($\mu=0,1,2$) with $\gamma_3$, so are proportional to either
$\mbox{tr}\gamma_\mu\gamma_3$ or $\mbox{tr}\gamma_5$. 

Now, for a theory with functional weight $\mbox{det}D_{Ls}[H]$ the corresponding
expression for $\langle\bar\psi\psi\rangle$ is
\begin{eqnarray}
\mbox{tr}M^{-1}M^\prime&=&\mbox{tr}[1-\gamma_3\varepsilon_{L_s}+m_h(1+\gamma_3\varepsilon_{L_s})]^{-1}
[1+\gamma_3\varepsilon_{L_s}]\nonumber\\
&=&\mbox{tr}{{[1+\gamma_3\varepsilon_{L_s}]}\over{[1-\gamma_3\varepsilon_{L_s}]}}\left[
1-m_h{{[1+\gamma_3\varepsilon_{L_s}]}\over{[1-\gamma_3\varepsilon_{L_s}]}}
+m_h^2{{[1+\gamma_3\varepsilon_{L_s}]^2}\over{[1-\gamma_3\varepsilon_{L_s}]^2}}+\cdots\right].
\label{eq:expansion}
\end{eqnarray}
Here $\varepsilon_{L_s}[H]\equiv\tanh(L_s\tanh^{-1}H)$ is the finite-$L_s$
approximation to the signum function. Now,
\begin{eqnarray}
{{1+\gamma_3\varepsilon_{L_s}}\over{1-\gamma_3\varepsilon_{L_s}}}&=&
[1-\gamma_3\varepsilon_{L_s}]^{-1}[1+\varepsilon_{L_s}\gamma_3]^{-1}[1+\varepsilon_{L_s}\gamma_3]
[1+\gamma_3\varepsilon_{L_s}]\nonumber\\
&=&(1-\varepsilon_{L_s}^2-[\gamma_3,\varepsilon_{L_s}])^{-1}
(1+\varepsilon_{L_s}^2+\{\gamma_3,\varepsilon_{L_s}\}).
\label{eq:dwprop}
\end{eqnarray}
In the limit $L_s\to\infty$, $\varepsilon_{L_s}^2=1$, and the long-wavelength
weak coupling limit (\ref{eq:wc}) gives
\begin{equation}
\lim_{L_s\to\infty}\{\gamma_3,\varepsilon_{L_s}\}=2;\;\;\
\lim_{L_s\to\infty}[\gamma_3,\varepsilon_{L_s}]=-{{2ip{\!\!\! /}}\over M},
\label{eq:limits}
\end{equation}
so (\ref{eq:dwprop}) $\approx 2M/ip{\!\!\! /}$ and we are on the right track.
However, for finite $L_s$ $1-\varepsilon_{L_s}^2$ is a real quantity, and now there
is no reason for the terms in (\ref{eq:expansion}) corresponding to
even powers of $m_h$ necessarily to vanish. Another way of saying this is that
the form of $A$ defining $\varepsilon_{L_s}$ 
dictates that it is no longer the case that even powers of $m_h$
are proportional to the trace over an odd number of gamma matrices. We conclude that the function
$\langle\bar\psi\psi(m_h)\rangle$ in general contains an even component, labelled
$\Delta_h$ in (\ref{eq:residuals}), weakly dependent on $m_h$ as $m_h\to0$ and only vanishing as $L_s\to\infty$.

Now repeat the exercise for the mass term $m_3S_3$:
\begin{eqnarray}
\mbox{tr}M^{-1}M^\prime&=&\mbox{tr}[1-\gamma_3\varepsilon_{L_s}+im_3\gamma_3(1+\varepsilon_{L_s}\gamma_3)]^{-1}
i\gamma_3[1+\varepsilon_{L_s}\gamma_3]\nonumber\\
&=&\mbox{tr}
\left[
1+im_3{{[1+\gamma_3\varepsilon_{L_s}]}\over{[1-\gamma_3\varepsilon_{L_s}]}}\gamma_3\right]^{-1}
{{[1+\gamma_3\varepsilon_{L_s}]}\over{[1-\gamma_3\varepsilon_{L_s}]}}
(i\gamma_3)
\label{eq:expansion3}
\end{eqnarray}
Now, from (\ref{eq:dwprop}) and the considerations of Sec.~\ref{sec:325},
all the terms in the binomial expansion of the first
factor in (\ref{eq:expansion3}) can only contain $L_s$ dependence in terms of the
form 
$(\gamma_3\varepsilon_{L_s})^p$, $\varepsilon_{L_s}^{2q}$ with $p,q$ integer,
which have the property that
$\mbox{tr}\gamma_3(\gamma_3\varepsilon_{L_s})^p=\mbox{tr}\gamma_3(\varepsilon_{L_s})^{2q}=0$.
This implies that only odd powers of $m_3$  
survive the trace. Hence
$i\langle\bar\psi\gamma_3\psi(m_3)\rangle$ is an odd function of $m_3$, and the
dominant residual $\Delta_h$ is necessarily absent. For finite $L_s$ when the
limiting forms (\ref{eq:limits}) do not hold, we cannot exclude corrections which
are odd functions of $m_3$, corresponding to the residual $\epsilon_3$ in
(\ref{eq:residuals}).

Finally, the arguments of Sec.~\ref{sec:325} then
imply the identical property for the condensate
$i\langle\bar\psi\gamma_5\psi\rangle$, consistent with the numerical results of
\cite{Hands:2015qha}.

\section{Summary}

In Sec.~\ref{sec:DW2ov} we showed that the 2+1$d$ domain wall fermion
formulation introduced in \cite{Hands:2015qha} coincides with the overlap
operator in the limit $L_s\to\infty$, and, importantly not simply in the
continuum limit as suggested in the abstract of that paper. 
Whilst the Dirac matrices $\gamma_3$ and $\gamma_5$ enter the domain wall
formulation (\ref{eq:SDWF}) in very different ways, it was shown in Sec.~\ref{sec:325} that
the resulting 2+1$d$ truncated overlap operator (\ref{eq:tanh},\ref{eq:tanh5}) is blind to the distinction
between them even for $L_s$ finite. There seems to be no
obstruction to modelling 
U($2N_f)\to$U($N_f)\otimes$U($N_f$) symmetry breaking in lattice simulations of
2+1$d$ fermions,
so long as it is understood that the nature of the $a>0$ corrections
to continuum symmetry operations, encapsulated in the GW relations
(\ref{eq:GW},\ref{eq:GW535}), and needed, say for identifying interpolating
operators for Goldstone modes~\cite{Hands:2015qha}, 
is more complicated than for 4$d$ gauge
theories, as discussed in Sec.~\ref{sec:GW}. 
In particular the antihermitian mass term (\ref{eq:m3S3}) consistent with
the GW relations contains an O($a$) correction of a form not present in the continuum
action. Ultimately, successful control of these corrections will depend on the locality properties of the
overlap operator $D_{ov}$~\cite{Hernandez:1998et}, which is a dynamical
question.

On the other hand, the freedom to formulate alternative mass terms in 2+1$d$
leads to a potentially important computational saving; as shown in
Sec.~\ref{sec:bilcon}, finite-$L_s$ corrections to bilinear condensates may be
classified by whether they are odd or even functions of the symmetry-breaking
mass $m_i$, and the dominant even component $\Delta_h$ is absent for the
antihermitian mass terms $S_3, S_5$, whose use in numerical simulations with finite
$L_s$ thus seems preferred, while recalling from Sec.~\ref{sec:m3m5} that the
correct formulation of the Pauli-Villars bulk correction
$\mbox{det}D_{L_s,L_s}^{-1}(1)$ requires the hermitian mass $1\times S_h$.

\section*{Acknowledgements}
This work was supported by a Royal Society Leverhulme Trust Senior Research
Fellowship, and in part by STFC grant ST/L000369/1. 
I continue to benefit enormously from  discussions with
Tony Kennedy.


\begin{thebibliography}{10}

\bibitem{Bernevig} B.A.~Bernevig and T.L.~Hughes, {\it Topological Insulators
and Topological Superconductors\/}, (Princeton University Press, 2013).

\bibitem{Son:2007ja}
  D.~T.~Son,
  Phys.\ Rev.\ B {\bf 75} (2007) 235423.

\bibitem{Janssen:2012pq}
  L.~Janssen and H.~Gies,
  Phys.\ Rev.\ D {\bf 86} (2012) 105007;\\
  F.~Gehring, H.~Gies and L.~Janssen,
  Phys.\ Rev.\ D {\bf 92} (2015) 8,  085046.

\bibitem{DelDebbio:1997dv}
  C.~Frick and J.~Jers\'ak,
  Phys.\ Rev.\ D {\bf 52} (1995) 340;\\
  L.~Del Debbio, S.~Hands and J.C.~Mehegan,
  Nucl.\ Phys.\ B {\bf 502} (1997) 269;\\
  S.~Christofi, S.~Hands and C.~Strouthos,
  Phys.\ Rev.\ D {\bf 75} (2007) 101701;\\
  S.~Chandrasekharan and A.~Li,
  Phys.\ Rev.\ Lett.\  {\bf 108} (2012) 140404;
  Phys.\ Rev.\ D {\bf 88} (2013) 021701;\\
  V.~Ayyar and S.~Chandrasekharan,
  Phys.\ Rev.\ D {\bf 91}, no. 6, 065035 (2015);\\
  S.~Catterall,
  arXiv:1510.04153 [hep-lat].

\bibitem{Schmidt:2015fps}
  D.~Schmidt, B.~Wellegehausen and A.~Wipf,
  arXiv:1511.00522 [hep-lat].

\bibitem{Burden:1986by}
  C.~Burden and A.N.~Burkitt,
  Europhys.\ Lett.\  {\bf 3} (1987) 545.

\bibitem{Hands:2015qha}
  S.~Hands,
  JHEP {\bf 1509} (2015) 047.

\bibitem{Ginsparg:1981bj}
  P.H.~Ginsparg and K.G.~Wilson,
  Phys.\ Rev.\ D {\bf 25} (1982) 2649.

\bibitem{Neuberger:1997fp}
  H.~Neuberger,
  Phys.\ Lett.\ B {\bf 417} (1998) 141;
  Phys.\ Lett.\ B {\bf 427} (1998) 353.

\bibitem{Kaplan:1992bt}
  D.B.~Kaplan,
  Phys.\ Lett.\ B {\bf 288} (1992) 342.

\bibitem{Neuberger:1997bg}
  H.~Neuberger,
  Phys.\ Rev.\ D {\bf 57} (1998) 5417.

\bibitem{Kennedy:2006ax}
  A.D.~Kennedy,
  {\it Algorithms for dynamical fermions\/},
  hep-lat/0607038.

\bibitem{Shamir:1993zy}
  Y.~Shamir,
  Nucl.\ Phys.\ B {\bf 406} (1993) 90.

\bibitem{Luscher:1998pqa}
  M.~L\"uscher,
  Phys.\ Lett.\ B {\bf 428} (1998) 342.

\bibitem{Hernandez:1998et}
  P.~Hernandez, K.~Jansen and M.~Luscher,
  Nucl.\ Phys.\ B {\bf 552} (1999) 363.

\end{thebibliography}
\end{document}